**Communicative Competence for Individuals**

**who require Augmentative and Alternative Communication:**

**A New Definition for a New Era of Communication?**

Janice Light & David McNaughton

The Pennsylvania State University

**Abstract**

In 1989, Light defined communicative competence for individuals with complex communication needs who require augmentative and alternative communication (AAC) as a dynamic, interpersonal construct based on functionality of communication, adequacy of communication, and sufficiency of knowledge, judgment, and skills. Specifically, Light argued that in order to demonstrate communicative competence, individuals who required AAC had to develop and integrate knowledge, judgment, and skills in four interrelated domains: linguistic, operational, social, and strategic. In 2003, Light expanded this definition and argued that the attainment of communicative competence is influenced not just by linguistic, operational, social, and strategic competencies, but also by a variety of psychosocial factors (e.g., motivation, attitude, confidence, resilience) as well as by barriers and supports in the environment. In the 25 years since this definition of communicative competence for individuals who use AAC was originally proposed, there have been significant changes in the AAC field. In this paper, we review the preliminary definition of communicative competence proposed 25 years ago, consider the changes in the field, and then revisit the proposed definition to determine if it is still relevant and valid for this new era of communication.

The silence of speechlessness is never golden. We all need to communicate and connect with each other – not just in one way, but in as many ways as possible. It is a basic human need, a basic human right. And more than this, it is a basic human power… (B. Williams, 2000; p. 248).

In this quote, Bob Williams, an expert communicator via augmentative and alternative communication (AAC), clearly articulates the singular importance of communication. Without access to effective communication, individuals with complex communication needs are consigned to live their lives with minimal means to express needs and wants, develop social relationships, and exchange information with others (Blackstone, Williams, & Wilkins, 2007). The ultimate goal of intervention for individuals with complex communication needs is to support the development of communicative competence so that these individuals have access to the power of communication – to interact with others, to have an influence on their environment, and to participate fully in society (Beukelman & Mirenda, 2013). Communicative competence is essential to the quality of life of individuals with complex communication needs for it provides the means to attain personal, educational, vocational, and social goals (Calculator, 2009; Lund & Light, 2007).

In 1989, Light proposed an initial definition of communicative competence as "…a relative and dynamic, interpersonal construct based on functionality of communication, adequacy of communication, and sufficiency of knowledge, judgment and skill in four interrelated domains: linguistic competence, operational competence, social competence, and strategic competence" (p. 137). In this paper, we consider this definition of communicative competence proposed 25 years ago, highlight the key changes in the AAC field over the past 25 years, and then revisit this definition of communicative competence to determine if it is still relevant and valid in today's fast-changing and dynamic era of communication.

## Preliminary Definition of Communicative Competence

The preliminary definition of communicative competence proposed by Light (1989) rests on three fundamental constructs: (a) functionality of communication, (b) adequacy of communication, and (c) sufficiency of knowledge, judgment and skill.

### Functionality of Communication

Historically, communication interventions focused on attempting to remediate speech and/or language impairments in isolation in an effort to "repair broken parts" (Lyon, 1998; p.204). These interventions seldom resulted in the attainment of functional communication skills for those with complex communication needs (e.g., Estrella, 2000; Fox & Fried-Oken, 1996). In order to ensure the attainment of communicative competence, AAC interventions need to focus not on the demonstration of isolated skills within labs, clinic rooms, or therapy sessions, but rather on actual communication performance within naturally occurring contexts (Light, 1989; Williams, Krezman, & McNaughton, 2008). The need for a focus on functional communication and participation within society is recognized in the World Health Organization's proposed International Classification of Functioning, Disability, and Health (Enderby, 2013; Simeonsson, Björck-Åkesson, & Lollar, 2012). A functional approach emphasizes functional outcomes in the real world, with intervention to build skills that have consequences that are valued by individuals with complex communication needs and their partners in daily life, including the ability to express needs and wants, exchange information, develop social closeness, and participate as required in social etiquette routines (Light, 1988).

The functionality of communication skills, that is, the success of the skills (or the lack thereof), depends on the communication demands present within the individual's environment, be it home, school, work, and/or the community. Martin Pistorius, an adult with a neurodegenerative condition who relies on AAC, highlighted the critical importance of functional communication skills to meet daily communication needs throughout the day:

> We need to look at every aspect of our lives, from the time we wake up in the morning, until we get up the following morning. We need to be able to communicate 24/7 like so-called "normal" speaking people do. (Pistorius, 2004; p. 3)

**Adequacy of Communication**

Hand in hand with the focus on the functionality of communication, consideration of communicative competence also requires a focus on the attainment of an adequate level of communication skills to meet environmental demands and reach communication goals (Light, 1989). The attainment of communicative competence does not require mastery of the art of communication; rather communicative competence is a threshold concept with a focus on the attainment of sufficient knowledge, judgment, and skills to meet communication goals and participate within key environments. An individual's communicative competence may vary across contexts depending on the partners, environments, and communication goals. For example, some individuals with complex communication needs may have developed adequate skills to meet the demands of interactions with familiar partners in routine contexts, but may struggle to communicate effectively with less familiar partners in more novel contexts where the demands are greater.

What defines adequacy of communication will vary depending on the goals of the individual who uses AAC and the communication requirements to meet those goals. Individuals who require AAC may define the success of intervention differently than professionals do, depending on their personal goals; these views must be respected. Wertz (1998) provided this account of the intervention that he planned for Doug who had aphasia following a stroke:

> Treatment ended before I thought it would. The progress Doug made in our two months together prompted me to urge continued treatment. I was more excited about Doug's progress than he was, and he was more satisfied with his progress than I was. About halfway through our second month, Doug indicated he was ready to go home. He had passed a driving test, qualified for disability income, and achieved sufficient communicative ability for his purposes. His plan was to become a person rather than a patient. That was his right, and he exercised it (p.31).

As described in this account, Doug determined that he had attained an adequate level of communication to meet his goals in his daily life; from his perspective, he had attained sufficient communicative competence for the situations that mattered most to him, and his priority was to return to living his life, rather than participating in further intervention.

**Sufficient Knowledge, Judgment, and Skills**

According to Light (1989), the adequacy of functioning required to attain communicative competence is predicated upon sufficient knowledge, judgment, and skills in four interrelated domains: linguistic, operational, social, and strategic. Linguistic and operational competencies reflect knowledge, judgment, and skills in the tools of communication whereas social and strategic competencies reflect knowledge, judgment, and skills in the use of these tools in daily interactions.

**Linguistic competence.** If individuals with complex communication needs are to develop communicative competence, they must develop sufficient knowledge, judgment, and skills in the linguistic code of the language(s) spoken and written in the individual's family and broader social community, including receptive skills and as many expressive skills in these languages as possible. In addition, they must also learn the language code of the AAC systems that they utilize, including the representational aspects of AAC symbols (Mineo Mollica, 2003) as well as the semantic and syntactic aspects required to express meaning (Blockberger & Sutton, 2003). Doing so is complicated by the fact that many AAC systems are not actually true language systems (Light, 1997). They are essentially semantic systems that include sets of symbols to convey concepts, but have no inherent syntax or morphology. Developing competence with the language code of the AAC systems is further complicated for there is an asymmetry (Smith & Grove, 2003) between the language code through which individuals who require AAC receive their input (i.e., the spoken language of their families and broader social community) and the language code through which they must express themselves (i.e., the form and content of multimodal expression that may include use of some speech or speech approximations, use of gestures or signs, and use of aided AAC symbols). Furthermore, individuals with complex communication needs typically have limited access to models of effective communication via AAC (Ballin, Balandin, Stancliffe, & Togher, 2011). Gus Estrella, an experienced and sophisticated communicator via AAC, emphasized the importance of concerted intervention to build the linguistic skills that underpin communicative competence:

Dig in, get the support of both the school and the social service agencies, get the devices funded, and make us work our little tails off until we master enough language to become competent communicators. (Estrella, 2000; p. 45).

**Operational competence.** Operational skills involve skills in the technical operation of AAC strategies and techniques, including: (a) skills to produce the hand or body positions, shapes, orientations, and movements for gestures, signs, or other forms of unaided communication (e.g., eye blink codes, head nod / shake); (b) skills to utilize selection technique(s) for aided AAC systems (e.g., direct selection with a finger or fist, eye gaze, scanning with a single switch); and, (c) skills to navigate and operate aided AAC systems accurately and efficiently (e.g., navigate between pages, enter codes to retrieve pre-stored vocabulary items). These operational skills must extend across the full range of modes used by the individual with complex communication needs, including both unaided and aided means of communication, both low tech and high tech (Beukelman, Fager, Ball, & Dietz, 2007; Hodge, 2007). Randy Horton described the significant demands of learning the operational skills for a single AAC system (approximately 96 hours in Randy's case) and the lack of instruction typically provided to support the development of these skills:

People without disabilities receive 12 years of writing and language teaching during school. I had next to none. …Usually the consumer is given 2 to 6 hours of teaching how to use the device. Extensive, intensive teaching during implementation is the key to success (Horton, Horton, & Meyers, 2001, p. 49)

**Social competence.** Individuals who require AAC must develop social competence to ensure appropriate functional use of AAC tools to meet their communication goals; they must learn when to communicate and when not, about what to communicate, with whom, when, where, and in what manner (Hymes, 1972). Social competence requires both sociolinguistic and sociorelational skills. Sociolinguistic skills refer to the pragmatic aspects of communication, in other words, discourse skills (e.g., taking turns, initiating and terminating interactions, maintaining and developing topics) and skills to express a wide range of communicative functions (e.g., requesting attention, requesting information, providing information, confirming). Sociorelational skills refer to the interpersonal aspects of communication that form the foundation for developing effective relationships. Light, Arnold, and Birmingham (2003) identified a range of sociorelational skills that may further the communicative competence of individuals who use

AAC (e.g., participating actively in interactions, demonstrating interest in partners, projecting a positive self image). Sociorelational skills bear special importance for individuals with complex communication needs who may face significant barriers to interpersonal relationships (Anderson, Balandin, & Clendon, 2011; Light et al., 2003). Jim Prentice, an expert communicator via AAC who worked as a statistical record keeper at a large company, poignantly illustrated the importance of developing social competence:

> When I started to work, I'm sure that all the employees surrounding my workstation probably thought that I was someone from Mars. I rode in on my motorized wheelchair and has some sort of device attached to my chair. I rode past them and they really didn't know whether I was able to talk. If they did talk to me, they weren't sure I was able to answer them. …I stopped them in their tracks, before they were frozen on the spot, and said, "Good morning, my name is Jim. How are all of you doing today?" Big smiles came on their faces, and they seemed to answer in unison, "We are fine, and it's nice to have you working with us." That sure broke the ice. I felt like one of the team then. I made sure I programmed a few jokes into my communicator so that it would make my conversations more friendly and comfortable for them. It worked! (Prentice, 2000; p. 209).

**Strategic competence.** Because of their significant disabilities, the substantial environmental barriers confronted in society, and the inherent restrictions of AAC systems, individuals with complex communication needs invariably confront limitations in their linguistic, operational, and/or social competence. In these cases, they must develop coping strategies to bypass these limitations and allow them to make the best of what they do know and can do (McNaughton et al., 2008; Todman, Alm, Higginbotham, & File, 2008; Williams, 2004). These compensatory strategies may be temporary, used for a time while the individual recovers or learns new linguistic, operational, and/or social skills; or the compensatory strategies may be required long term in situations where limitations in the linguistic, operational, and/or social domain cannot be remediated (Light, 1989). In order to obtain communicative competence, individuals with complex communication needs may rely on a range of strategies to overcome linguistic constraints (e.g., asking the communication partner to write or type as they speak to support comprehension difficulties; directing the partner to provide choices when faced with vocabulary limitations); operational constraints (e.g., using telegraphic messages to enhance the rate of communication; asking partners to guess as messages are spelled to reduce fatigue); and social constraints (e.g., using an introduction strategy to explain the AAC system and how to use it; using humor to put unfamiliar partners at ease) (Miranda & Bopp, 2003). Randy Kitch, an expert communicator who uses his foot to access AAC, illustrated the importance of strategic competence to overcome the difficulties that he encountered when a store clerk ignored his communicative attempts:

> I decided to type him a note explaining how I communicated with my letter board and went back to the store the next day to give it to him. I went up to him, sat on the floor and footed him the note. It said, "I communicate by spelling words on a letter board with my big toe and I would appreciate it if you would communicate with me." It also said, "I would like to purchase some head cleaner for my cassette player." He got the cleaner. I gave him the money, and after he handed me the cleaner, I spelled out "THANK YOU" on the letter board and he said, "You're welcome." (Kitch, 2005; p.49).

## Psychosocial Factors that Influence Communicative Competence

In 2003, Light expanded the preliminary model of communicative competence and argued that the attainment of communicative competence by individuals with complex communication needs is impacted not just by their linguistic, operational, social, and strategic competence, but also by a range of psychosocial factors including motivation, attitude, confidence, and resilience.

**Motivation.** Motivation to communicate impacts the individual's desire or drive to communicate with others in daily situations (Light, 2003). Communication via AAC is a complex process that imposes significant motor, cognitive, sensory perceptual, and linguistic demands (e.g., Thistle & Wilkinson, 2013). When motivation to communicate is high, individuals with complex communication needs are more likely to tackle the demands of communication via AAC; when motivation is low, they may be overwhelmed by these demands and may elect to forego many communication opportunities (Clarke, McConachie, Price, & Wood, 2001; Fox & Sohlberg, 2000). Jan Staehely (2000), who utilizes AAC to support her communication, described the challenge of maintaining motivation when she did not have effective means to communicate:

> I had become so used to not being able to say something in depth to a person that I started to believe that I was a person who didn't have much to tell people. … I fooled myself into thinking that I didn't have anything to say. (p. 9).

Individuals with complex communication needs require numerous positive and successful communication experiences to build their motivation to attain communicative competence.

**Attitude.** The attitudes of individuals with complex communication needs and their families, especially as they relate to AAC also impacts the attainment of communicative competence. Attitudes towards AAC may predispose the use (or lack of use) of AAC as required within social situations. Lasker and Bedrosian (2000) proposed a model of AAC acceptance that considered the impact of three sets of factors: (a) milieu factors (e.g., partners, setting, time of day); (b) person factors (e.g., disability, personality, age, skills, needs, history, expectations); and (c) AAC-related factors (e.g., ease of learning, appearance, functionality). Attitudes may change with changes to person, milieu, and /or system factors. Rob Rummel-Hudson the father of a daughter, Schuyler, who requires AAC, described the effect of different AAC systems on his daughter's attitude toward AAC and, as a result, her willingness or unwillingness to utilize AAC to support her communication:

> Her enthusiasm [with her new SGD] was perhaps the most significant development, perhaps more important than whether or not she intuitively "got it." She did, but even better, she was fascinated by the device. She used it for everything. …We knew that if a speech prosthesis was going to work for her, it was going to be because she took the initiative to make it happen, the same way she came to embrace sign language and, conversely, the way she completely rejected the picture identification system that every one of her schools had tried to get her to use… My pity went out to the person who tried to make Schuyler do something she didn't want to do, or who tried to keep her from doing something she liked. (Rummel-Hudson, 2008; p. 223).

**Confidence.** Motivation impacts the individual's drive to communicate and attitude toward AAC impacts the individual's willingness to use AAC to communicate, but it is confidence that actually determines the individual's propensity to act – in other words, to attempt to communicate in any given situation. Confidence has to do with the individual's self-assurance, in this case, specifically self-assurance that he or she can communicate successfully in the given situation(s). Using AAC to meet communication needs requires individuals with complex communication needs to try techniques that may initially be new and unfamiliar, to both them and their communication partners, typically with few models of others who successfully communicate using AAC  (Ballin et al., 2011; Light et al., 2007). Seeing or interacting with others who use AAC who have attained communicative competence may serve as a critical support in building communicative confidence. Rick Creech (1995), who was a pioneer in his use of

AAC in both post-secondary settings and the workplace, explained,

> Until we have seen a fluent interactive, augmented speaker who shares our physical circumstances, there may have been little in our personal experience to indicate that we, ourselves, would someday actually 'talk'. (p. 12).

**Resilience.** Although confidence may determine the individual's propensity to attempt to communicate, it is resilience that influences whether or not the individual perseveres with communication despite the many challenges and potential failures encountered. Resilience refers to the "…capacity which allows a person … to prevent, minimize, or overcome the damaging effects of adversity" (Grotberg, 1995, p. 7). It is inevitable that individuals with complex communication needs will confront failure at times in their attempts to communicate successfully. These failures may result from limitations in their linguistic, operational, social, and strategic skills and/or from barriers within the environment (Balandin, Hemsley, Sigafoos, & Green, 2007; Snell, Chen, Allaire, & Park, 2008). Communication failures provide important opportunities for learning and may ultimately fuel subsequent success, but only if the individual is resilient enough to move on and try again.

Resilience is a dynamic factor that is affected by the adversity encountered (e.g., the nature, severity, timing of the adversity), as well as protective factors (both individual resources and environmental supports) that may support recovery from the adversity (Luthar, Cicchetti, & Becker, 2000; Masten, 2001). For example, individual protective factors that support resilience may include problem solving skills, self esteem, optimism, or faith; environmental protective factors may include encouragement and support from family, mentors, teachers, employers, or peers. Individuals with complex communication needs who have access to clusters of protective factors are more apt to demonstrate resilience in the face of communication failures and are therefore more apt to build, re-build, or sustain communicative competence in the face of adversity (Dickerson, Stone, Panchura, & Usiak, 2002; Smith & Murray, 2011). In contrast, those who do not have access to protective factors will have greater difficulty rebounding from adversity, learning from these failures, and ultimately developing communicative competence. Morrie Schwartz, a man who had ALS, wrote about the importance of resilience in the face of the adversity that he faced as his disease progressed:

> I have become more and more dependent as my disease has progressed. I am being wheeled around to get everywhere, I am fed, bathed, taken to the john. A whole host of things I did independently and took for granted as being part of my physical self are now done for me by other people. Although I am dependent, I have an independent mind, mature emotions, and I use my independence to keep my essential self going. (Pillar & Schwartz, 1996; p. 73).

**Environmental Supports and Barriers**

Communicative competence is impacted not only by factors intrinsic to the individual with complex communication needs (e.g., linguistic, operational, social and strategic skills as well as psychosocial factors such as motivation, attitude, confidence and resilience), but also by extrinsic factors, including barriers in the environment that may impede communicative competence, and environmental supports that may enhance communicative competence (Light, 2003). Ultimately, communication is an interpersonal process where meaning is created in partnership (Blackstone et al., 2007; Teachman & Gibson, 2014). As a result, intervention to enhance communicative competence necessitates not only intervention with the individual with complex communication needs, but also intervention with partners in the environment to reduce barriers and ensure appropriate supports as required (Ball & Lasker, 2013; Kent-Walsh & McNaughton, 2005; Soto, 2012). Jan Staehely, who uses AAC to

communicate, emphasized the interpersonal nature of communicative competence as follows:

Just as a dance couldn't possibly be a dance unless people moved to it, so language doesn't become communication until people grow to understand and express it back. It has to be a two-way exchange. This is why communicating is an action word. (Staehely, 2000; p. 3).

All individuals who require AAC are impacted by environmental factors, but the extent of the impact will vary across individuals depending on their intrinsic communication resources: Those with strong linguistic, operational, social, and strategic skills and well-developed psychosocial factors will be less vulnerable to environmental barriers and constraints than those who are beginning communicators or those who experience significant language /cognitive limitations (McNaughton & Light, 2013; Williams, Beukelman, & Ullman, 2012) According to Beukelman and Mirenda (2013), environmental barriers and supports may cut across a range of domains including policy, practice, attitude, knowledge and skill barriers or supports.

**Policy and practice barriers and supports.** Individuals with complex communication needs may encounter policy and practice barriers that are systemic within society and serve to limit their communication opportunities and therefore their development of communicative competence (Cooper, Balandin, & Trembath, 2009; Stancliffe et al., 2010). Policy barriers result from official laws, standards, or regulations, whereas practice barriers result from conventional procedures within schools, work settings, or society that may not be officially documented but are accepted practice (Beukelman & Mirenda, 2013). John Draper, a competent communicator who relies on AAC, discussed some of the policy / practice barriers he encountered during his education in an inclusive school environment:

> My success in meeting the rigors of the school curriculum depended in large part on the extent to which my educational team worked collaboratively. It was not uncommon for more than 30 professionals to be involved in my life at any one time. It was a constant struggle to get everyone to work together effectively and not to become distracted by their individual mandates, policies, and turfs. It took time for everyone to realize that true collaboration could be achieved only when the team understood everyone's individual roles, clarified expectations in writing, and established communication guidelines. (Carter & Draper, 2010; p. 73).

Sometimes practices that appear to be inconsequential to professionals have substantial negative effects on the lives of individuals who require AAC. John Draper described some of the practices at his high school that created barriers in his interactions with his peers:

> Of utmost importance to me was having a sense of belonging in my school community. By virtue of my physical and communication challenges, I didn't really fit into the social circles of high school. This reality, combined with the lack of knowledge on the part of many school personnel on how to promote disability awareness or foster peer relationships, resulted in missed opportunities. One example in high school was how lockers of students who had a disability were grouped in a separate location rather than integrated into the alphabetical order of the rest of the student population. Another example was the practice of having students who had a disability work with paraprofessionals in a segregated resource room during free periods rather than allowing us to interact with our peers in the school library. These practices limited my chances of connecting with my peers. (Carter & Draper, 2010; p. 82).

Ultimately, as Carter and Draper (2010) suggest, the goal is to eliminate policy and practice barriers and ensure that there are sufficient supports for participation and meaningful inclusion of individuals who require AAC in all aspects of society. Mirenda (1993) summed up this goal best when she wrote:

> I am talking about community living situations that help people become members of, not just residents in, communities. I am talking about programs in which a lot of emphasis is placed on helping people get to know and connect with their neighbors and their local shopkeepers. …(M)embership is different than joining or living next door to or affiliating with  - you can do all those things on your own. But you have not achieved membership in a group until the group says you have; it is mutual, it is consensual. That is what we want – membership in communities. (p. 6)

**Attitude barriers and supports.** As Mirenda (1993) suggested, achieving true membership in communities is not just about policy and practice supports, it also requires the elimination of attitude barriers. Attitude barriers occur when people hold negative feelings that predispose them to act in ways that limit the communication opportunities of individuals who require AAC (Hodge, 2007; McCarthy & Light, 2005). Bob Williams (2000) described the problem of pervasive attitude barriers for individuals who require AAC:

> Why are so many people consigned to lead lives of needless dependence and silence? Not because we lack the funds, nor because we lack the federal policy mandates needed to gain access to those funds. Rather, many people lead lives of silence because many others still find it difficult to believe that people with speech disabilities like my own have anything to say or contributions to make. (p. 250).

As Williams suggested, too often attitude barriers result in reduced expectations for individuals with complex communication needs and limited opportunities for participation. Concerted advocacy and intervention is required to address attitude barriers and ensure that individuals with complex communication needs who use AAC have meaningful opportunities to communicate and to participate at school, at work, in their families, and in their broader social communities.

**Knowledge and skill barriers and supports.** Even when the necessary policy, practice and attitude supports are in place, it may not be sufficient to ensure the development of communicative competence by individuals who require AAC. Learning to communicate using AAC is a complex process (Bailey, Parette, Stoner, Angell, & Carroll, 2006; Rackensperger, Krezman, McNaughton, Williams, & D'Silva, 2005). Many individuals who require AAC experience significant linguistic, operational, and social constraints and require support from their partners to ensure successful communication (Blackstone et al., 2007). In order to provide appropriate supports, partners (e.g., family members, educational personnel, employers, co-workers, friends) require knowledge of AAC systems and services as well as competencies in appropriate interaction strategies (e.g., Kent-Walsh, Binger, & Hasham, 2010; Sorin-Peters, McGilton, & Rochon, 2010). Jean Dominique Bauby (1997) emphasized the importance of the partner's knowledge and skill in determining the success (or failure) of his communication attempts using AAC following a brainstem stroke:

> It is a simple enough (AAC) system. You read off the alphabet …until, with a blink of my eye, I stop you at the letter to be noted. …That, at least, is the theory. In reality, all does not go well for some visitors. Because of nervousness, impatience, and obtuseness, performances vary in the handling of the code. …Nervous visitors come most quickly to grief. They reel off the alphabet tonelessly, at top speed, jotting down letters almost at random; and then seeing the meaningless result, exclaim, "I'm an idiot!". …Reticent people are much more difficult. If I ask them, "How are you?" they answer, "Fine," immediately putting the ball back in my court. …Meticulous people never go wrong: they scrupulously note down each letter and never seek to unravel the mystery of a sentence before it is complete. …Such scrupulousness makes for laborious progress, but at least you avoid the misunderstandings in which impulsive visitors bog down when they neglect to verify their intuitions. (p. 20-22).

As Bauby (1997) suggested, partners may require instruction to develop the knowledge and skills required to interact effectively and support communicative competence with individuals who require AAC.

**Key Changes in the Field of Augmentative and Alternative Communication**

In the 25 years since Light first proposed this model of communicative competence, there have been dramatic changes in the AAC field: (a) changes in the demographics of the population that uses AAC; (b) changes in the scope of communication needs that must be considered; (c) changes in the AAC systems that are available; and, (d) changes in expectations for participation by individuals who use AAC (Light & McNaughton, 2012a). Given these dramatic changes, it seems appropriate to re-visit the original definition of communicative competence to assess its current relevance and validity. Specifically, we consider each of the key changes in the field and the potential implications of these changes for the proposed model of communicative competence as well as the implications for interventions to build communicative competence.

## Changes in the Demographics of the Population that Uses AAC

During the past 25 years, the field of AAC has witnessed significant increases in the numbers of people with complex communication who receive or might benefit from AAC services; furthermore, the population receiving AAC services is increasingly diverse in terms of age, disability, language, culture, and race/ ethnicity (Beukelman, 2012; Light & McNaughton, 2012a; Mueller, Singer, & Carranza, 2006; Soto & Yu, 2014). In addition to the increased prevalence of individuals with complex communication needs, there have also been significant improvements in preservice and inservice training in AAC over the past 25 years (e.g., Costigan & Light, 2010; Ratcliff, Koul, & Lloyd, 2008), resulting in greater professional awareness and acceptance of AAC intervention generally. AAC interventions are no longer viewed only as a last resort to be implemented with individuals with no speech or extremely limited speech, only once traditional speech and language interventions fail; rather an increasing number of professionals now understand the potential benefits of AAC intervention for those who are at risk for speech and language development (e.g. Romski et al., 2010), those who rely on speech but require augmentation to clarify or enhance intelligibility (e.g., Hanson, Beukelman, & Yorkston, 2013), those who are recovering following a stroke or traumatic brain injury (e.g., Petroi, Koul, & Corwin, 2014), those who are experiencing the loss of speech or language skills due to degenerative conditions (e.g., Fried-Oken, Beukelman, & Hux, 2012) and those who may have temporary conditions (e.g., Costello, Patak, & Pritchard, 2010). As a result, AAC interventions are now implemented with a much larger and more diverse population, including individuals across the life span, both younger and older than ever before, and individuals with a wide array of disabilities who present with a much more diverse array of needs and skills than ever before.

Beyond the increased diversity in the age and disability profiles of individuals who require AAC, there is also increased diversity in language, culture, and ethnicity/ race of those who are receiving AAC services (Soto & Yu, 2014). This linguistic, cultural, and racial/ ethnic diversity results from two key developments. First, the global reach of AAC intervention has been extended worldwide, especially to developing countries, through the efforts of families and professionals (Alant, 2007; Bornman, Bryen, Kershaw, & Ledwaba, 2011). Evidence of the growing impact of AAC worldwide is found in the recognition of the International Society for Augmentative and Alternative Communication (ISAAC) as a Non-Governmental Organization in consultative status with the United Nations Economic and Social Council. In addition to this extended global reach of AAC, Soto and Yu (2014) noted that unprecedented movement of the population over the past 20-25 years (e.g., from developing countries to developed ones, from rural to urban areas) has resulted in substantial increases in the numbers of children and adults with complex communication needs receiving AAC services who come from culturally and linguistically diverse backgrounds.

**Implications of the changing demographics for communicative competence.** What are the implications of these changing demographics for the definition of communicative competence and for interventions to enhance communicative competence? The greater range of ages and disabilities served has necessitated a greater range of AAC interventions, including ones: (a) to build communicative competence for the first time with those who have developmental disabilities through instruction in linguistic, operational, social, and strategic skills (e.g., Snell et al., 2010); (b) to re-build communicative competence with those who have acquired disabilities or temporary conditions, capitalizing on existing linguistic and social strengths and teaching operational and strategic skills to bypass limitations in these domains to maximize communication performance (e.g., Costello et al., 2010, Petroi, at al., 2014; Light & Gulens, 2000; Simmons-Mackie, King & Beukelman, 2013); and, (c) to sustain communicative competence for as long as possible with those who have degenerative neurogenic disabilities through implementation of AAC supports (e.g., Fried-Oken et al., 2012).

These interventions must respond not only to the motor, sensory perceptual, and cognitive skills of individuals who require AAC, but also to their cultural and linguistic backgrounds (Binger, Kent-Walsh, Berens, Del Campo, & Rivera, 2008). Individuals with complex communication needs who live in bilingual or multilingual environments face significantly increased linguistic and operational demands in the development of communicative competence, for the different languages and cultures in which they participate will no doubt require different modalities of communication, different vocabularies, different representations, different layouts, and different organizations (e.g., Nakamura, Iwabuchi, & Alm, 2006; Soto & Yu, 2014). Individuals who require AAC who live in bilingual and multilingual environments must develop competence in: (a) the spoken and written languages of their family and broader social communities including comprehension skills and as many expressive skills as possible, including the phonological, semantic, syntactic, morphological and pragmatic aspects of these languages, which may differ significantly depending on the specific languages involved; (b) the language codes of the different AAC strategies and techniques that they use to communicate in these different cultural and linguistic environments; (c) the operational skills to produce and /or technically operate these different unaided or aided AAC systems; and (d) the social skills to know when and how to code switch between languages and different AAC strategies and techniques across different environments. Clearly the linguistic, operational, and social demands to attain communicative competence are multiplied when individuals with complex communication needs come from bilingual or multilingual environments. Estrella (2000) poignantly described the challenges:

> Prior to starting preschool, my family and friends all spoke to me in Spanish. That was all I knew. So you can imagine my reaction when I started going to preschool. I was entering uncharted territories. I was about to be left with total strangers, foreigners! It was doubtful that anyone would know any Spanish, so what was the likelihood of somebody understanding my little signs for when I needed something, like lunch! What if I need to go to the little boys' room and they think I'm having a seizure! These were the concerns that a little boy had to deal with and figure out how to cope with his new surroundings. …I felt isolated since I couldn't tell anybody what I was thinking or feeling. (p. 33).

Soto and Yu (2014) highlighted the benefits of bilingual intervention for individuals with communication disabilities. However, they noted that in order to provide effective bilingual intervention, AAC professionals must develop the competencies required to provide culturally competent services, specifically the skills to: (a) accurately assess communication skills of individuals with complex communication needs who come from diverse cultural and linguistic backgrounds; (b) support language development and/or recovery across the languages of the family and broader social community; (c) select, customize, and implement culturally appropriate AAC strategies and techniques to support communication across diverse environments; and (d) work effectively with families from diverse backgrounds. The increased diversity of the population that would benefit from AAC, in terms of age, disability, language, and culture has increased the urgent need for high quality preservice and inservice training to ensure that professionals from multiple disciplines have the competencies required to provide effective, culturally-competent, evidence-based AAC services to foster communicative competence with individuals across the life span, who present with a wide array of needs and skills (Costigan & Light, 2010; Soto & Yu, 2014).

**Changes in the Scope of Communication Needs**

Along with the changes in the demographics of the population that requires AAC have come dramatic changes in the scope of the communication needs that must be addressed. Twenty five years ago, there was an emphasis on providing the means to express needs and wants; increasingly there is a growing recognition that communication extends well beyond needs and wants, and must serve to foster the development of social relationships, the exchange of information and participation in social etiquette routines (De Leo, Lubas, & Mitchell, 2012; Waller et al., 2013). Perhaps the mother of Brian, an 8-year-old boy with severe multiple disabilities, summed it up best when she said, "There's more to life than cookies." (Light, Parsons & Drager, 2002; p. 187). In fact, with the advent of social media and a new arsenal of tools to link people together, there is increased emphasis in society on establishing, maintaining, and developing social connections across a wide ranging network (Sundqvist & Ronnberg, 2010; Williams et al., 2012).

Twenty-five years ago, the focus was primarily on maximizing the communication of individuals with complex communication needs within face to face interactions. Now there is increased recognition that communication needs extend well beyond face to face interactions and also include written communication to meet demands at school or in the work place; social media such as Facebook and Instagram to network, share experiences, and establish membership in peer communities; cell phone and texting to connect with friends; blogging to provide commentary and build communities with like interests; Twitter to instantaneously update status and express short bursts of opinion; e-commerce to fulfill a wide array of needs and wants, and so on (Light & McNaughton, 2012a).

**Implications of the changing scope of communication for communicative competence.** With the dramatic change in the scope of communication and the explosion of tools through which to meet communication needs, individuals with complex communication needs now have access potentially to a much wider and more diverse audience than ever before. They have mechanisms available to address what was previously one of the greatest barriers – that of limited social networks and communication partners (Blackstone & Hunt Berg, 2003). Glenda Watson Hyatt (2011) who uses a variety of AAC technologies (including the iPad) to communicate described the deeper level of communication possible as a result of the greater range of social media tools:

The cool thing was … I had Internet access. When asked what I had been up to, I responded 'problogging and ghost writing,' and I was able to show what I had written. I also shared the video of me ziplining across Robson Square in downtown Vancouver during the Winter Olympics. The iPad allowed for a deeper level of communication than would have been possible with a single-function AAC device. (p.25)

With access to an increased array of potential partners, however, have come increased demands for independent and easily intelligible communication. In using these media tools, individuals with complex communication needs cannot co-construct messages with familiar partners as they do in face to face interactions; rather they must develop the skills to independently use these new tools, adhere to their conventions, and communicate with a broader audience including those who may have limited or no prior experience with AAC. In general terms, establishing greater independence and intelligibility of communication to reach a wider audience requires more advanced linguistic skills, specifically the ability to effectively convey meaning through traditional orthography with appropriate form and content as required by the target media and audience (Fager, Bardach, Russell, & Higginbotham, 2012).

Interestingly, many of these new social media do not rely solely on linguistic content to communicate; rather linguistic content may be supplemented with extensive use of visual images (i.e., photos, video) as channels of expression. This trend towards increased use of photos and video has some potential advantages for individuals with complex communication needs for use of visual images such as photos to enhance communication has a long history in the AAC field (Hanson, et al., 2013). With the advent of many social media applications, photo and video have become widely-accepted channels of expression across society (Light & McNaughton, 2012a), and are used to support communication for educational, employment, health, and social purposes (Raghavendra, Newman, Grace, & Wood, 2013).

However, in order to effectively use these diverse media to enhance communication on social media platforms (e.g., Facebook, Twitter, Instagram,), individuals with complex communication needs typically require functional literacy skills as well as the ability to capture photo and video of meaningful events and experiences. Thus, these media impose increased linguistic demands (e.g., functional literacy skills; semantic, syntactic and morphological skills) and increased operational demands (e.g., capture and posting of photo and video). Furthermore, for each communication media, individuals with complex communication needs must learn the rules of social use (i.e., with whom to communicate, about what, when, where, in what form, and for what purposes).

These rules vary across media: For example, written papers for school or reports for work require formal vocabulary, syntax, and morphology, whereas Twitter is limited to 140 characters, with the use of sentence fragments and spelling abbreviations acceptable to provide status updates and express opinions. Furthermore, individuals who use AAC must learn the sociolinguistic rules for using each of these media without the benefit of immediate, visible, partner feedback. Given the dominance of social media in today's society and the potential benefits for individuals with complex communication needs, future research is required to investigate the use of social media and other mainstream communication tools by individuals who require AAC.

## Changes in AAC Systems

Implicit in considering the dramatic changes to the scope of communication needs is the realization that individuals with complex communication needs can no longer rely on a single speech generating device to meet their communication needs if they are to participate fully within educational, vocational and social contexts (Williams et al., 2008). Rather they must have access to a wide range of means to augment and enhance their communication that may include unaided AAC (e.g., gestures, signs, speech or speech approximations), low tech aided AAC systems (e.g., communication boards or books), high tech AAC systems (e.g., traditional speech generating devices, mobile technologies with AAC apps), and other mainstream communication apps and social media tools (e.g., Facebook, Twitter, Instagram, SnapChat).

**Implications of changes in AAC systems for communicative competence.** The dramatic changes in the range of AAC systems/ apps, communication technologies, and social media tools bring both benefits and challenges in terms of building, rebuilding, and sustaining the communicative competence of individuals who require AAC. The iPad and mobile technology revolution and the greater use of social media tools have positively impacted social awareness and acceptance of AAC, reducing attitudinal barriers to AAC use (McNaughton & Light, 2013). Individuals with complex communication needs may be more apt to make use of these tools as AAC techniques to enhance communicative competence as a result. Rob Rummel-Hudson, a parent of a teenager who uses AAC, emphasized the positive effects of mobile technologies on attitudes of individuals with complex communication needs and their families:

> …[the iPad] provides a rather elegant solution to the social integration problem. Kids with even the most advanced dedicated speech device are still carrying around something that tells the world 'I have a disability.' Kids using an iPad have a device that says, 'I'm cool.' And being cool, being like anyone else, means more to them than it does to any of us. (Rummel-Hudson, 2011; p.22)

Although there are substantial benefits to the increased range of AAC systems/ apps, social media, and mainstream communication tools that are available to individuals who require AAC, there are also significant challenges. The increased diversity of communication tools means substantially increased operational demands for individuals with complex communication needs. Each of the tools is designed with different representations, organizations, and layouts of information as well as different access techniques (e.g., swiping, tapping, double tapping). And each of these different designs imposes different motor, cognitive, sensory perceptual and linguistic learning demands for individuals with complex communication needs. Typically these tools are not well integrated, increasing the operational demands on individuals with complex communication needs who must not only learn operational skills for each tool, but also acquire the skills to navigate between apps or tools as required.

The development of operational competence lies at the intersection of the demands imposed by the communication technologies and the intrinsic skills of the individual who requires AAC.

Traditionally the focus of intervention has been on teaching individuals with complex communication needs the necessary motor skills; however, research demonstrates that visual, cognitive, and linguistic processing skills also play critical roles in determining operational competence (e.g., Costigan, Light, & Newell, 2012; Wilkinson, Light & Drager, 2012).

To date, most attention has focused on intervention to teach skills to the individual with complex communication needs. Much less attention has been directed towards improving the design of AAC systems specifically and the design of mainstream social media tools generally to reduce operational demands, ease learning, and facilitate use. As Light and McNaughton (2012b) noted, "The lack of attention to the design of AAC technologies/apps is rather ironic since this component of intervention is one that substantially affects performance and it is also the one that is most easily amenable to change." (p. 36). Clearly future research is required to investigate the basic visual, cognitive, linguistic and motor processing demands of AAC systems and to untangle the effects of specific system components in order to optimize the designs of AAC systems and social media tools, and thus support operational competence for individuals with a wide range of disabilities.

There is an urgent need to define basic design specifications to facilitate use across apps and social media tools for people with disabilities, and to support rapid individualization that will provide access to persons with specific disabilities and strengths (Emiliani, Stephanidis, & Vanderheiden, 2011; Vanderheiden et al., 2012).Without these principles in place, individuals with complex communication needs are forever playing catch up, trying to learn new operational requirements as new technologies emerge, or they are excluded from access to apps and social media tools when the designs impose demands outside of their motor, sensory perceptual, linguistic and cognitive capabilities. With increased diversity in the scope of communication needs and increased availability of a wide range of AAC systems and social media tools to meet these needs, there is even greater need than ever before for the input of multidisciplinary teams with expertise in a wide range of domains extending well beyond expertise in traditional speech and language skills to include expertise in literacy skills, human computer interface, visual cognitive processing, motor performance, and instructional design, to name just a few. No longer can AAC intervention be limited in focus to the use of speech generating devices in face to face interactions; rather intervention must extend well beyond speech prostheses to maximize communication across a broad array of media (Shane, Blackstone, Vanderheiden, Williams, &

DeRuyter, 2012). Concerted advocacy is required to ensure that public policy and funding agencies keep pace with these developments; they must recognize and support access to the wide breadth of communication tools required for full participation in educational, vocational, and social contexts (Vanderheiden et al., 2013).

**Changes in Expectations for Participation**

Twenty five years ago, many individuals with complex communication needs lived in large residential institutions, segregated from their families and communities with limited educational and vocational options (Mirenda, 2014). Now, however, increasing numbers of individuals with complex communication needs live within their communities (Lakin & Stancliffe, 2007); attend schools with the other children in the neighborhood and participate in general education classrooms (e.g., Calculator, 2009); obtain full time or part time work through community jobs, telework, or micro-enterprises (e.g., Isakson, Burgstahler, & Arnold, 2006; McNaughton, Rackensperger, Dorn, & Wilson, in press); and engage in a wide range of meaningful activities within the community (Thirumanickam, Raghavendra, & Olsson, 2011; Trembath, Balandin, Stancliffe, & Togher, 2010).

**Implications of changing participation patterns for communicative competence.** With these changes in living, schooling, employment and community living have come substantial increases in the communication demands for individuals with complex communication needs across these different environments (Johnson, Douglas, Bigby, & Iacono, 2009; Raghavendra, Virgo, Olsson, Connell, & Lane, 2011). Twenty five years ago, many individuals with complex communication needs only had the opportunity to interact with the staff in the institutions and residences in which they lived; they were pre-empted from many communication opportunities and had only limited choices. Now, individuals with complex communication needs require AAC systems to support their communication and full participation at home, at school, at work, in health care settings, and within the community (Collier, Blackstone, & Taylor, 2012; Collier & Self, 2010). It is no longer sufficient for individuals with complex communication needs to have access to the means to simply request a preferred food or activity; rather they need access to communication to build friendships with peers, to learn at school, to share their expertise on the job, to manage their health care needs, and to participate successfully as full citizens of society (Bryen, Chung, & Lever, 2010; Kennedy, 2010). Individuals with complex communication needs face increased requirements for linguistic, operational, social, and strategic competencies to meet the increased communication demands of participation in diverse environments (e.g., home, school, work, community). AAC interventions must serve to help build the necessary knowledge, judgment, and skills to ensure the development of communicative competence. With increased expectations for full participation in society, individuals who require AAC now interact with a much broader range of partners in much more diverse contexts than ever before and thus face increased communication demands on a daily basis as a result. In order to meet these challenges, it is more critical than ever for individuals with complex communication needs to develop the necessary protective factors to fortify their motivation, attitude, confidence, and resilience in the face of the adversity that they will no doubt face at times. Furthermore, there is increased need for intervention to break down environmental barriers in society that limit participation and to replace them with positive supports to enhance the communicative competence of individuals who require AAC (Johnson et al., 2009).

**Research to Advance Understanding of Communicative Competence**

Over the past 25 years, there has been a significant increase in research to advance understanding and enhance the communicative competence of individuals with complex communication needs. This research has established empirical evidence of the positive impact of AAC (Beukelman et al., 2007; Bopp, Brown, & Mirenda, 2004; Branson & Demchak, 2009; Fried-Oken et al., 2012; Ganz et al., 2011; Machalicek et al., 2010;

Roche, et al., 2014; Schlosser, Sigafoos, & Koul, 2009; Walker & Snell, 2013; Wendt, 2009) and has demonstrated that these gains come at no risk to speech development or recovery (e.g., Millar, Light, & Schlosser, 2006; Romski et al., 2010). As a field, we should take pride in this increased research base that has resulted in advances in evidence-based AAC services. Over the past 25 years, we have also witnessed increased involvement of individuals with complex communication needs and their families in these research endeavors, working to ensure that their voices are heard and their needs and priorities are addressed (O'Keefe, Kozak & Schuller, 2007; Rackensperger et al., 2005)

Despite these significant advances, there remain many unanswered questions regarding effective interventions to build, rebuild, or sustain communicative competence with the diverse range of individuals across the life span who have developmental, acquired, degenerative, and temporary disabilities resulting in complex communication needs. Future research is required to investigate effective interventions: (a) to enhance the knowledge, judgment, and skills of individuals with complex communication needs across all domains - linguistic, operational, social, and strategic; (b) to fortify psychosocial supports to maximize motivation, positive attitudes, confidence, and resilience; and (c) to eradicate environmental barriers (i.e., policy, practice, attitude, knowledge and skill barriers) and ensure appropriate supports from communication partners in home, school and community environments to further the communicative competence of individuals with complex communication needs.

**Conclusions**

In conclusion, it is clear that the definition of communicative competence for individuals who require AAC, first proposed by Light 25 years ago (1989), continues to provide a useful framework for this new era of communication. Despite the dramatic changes in the demographics of the population that requires AAC, the scope of communication needs to be addressed, the range of AAC systems/ apps and social media tools available, and the expectations for participation across a wide range of environments, the essential goal of intervention has not changed. AAC interventions must address the development of adequate, functional communication skills to support individuals with complex communication needs in developing, rebuilding, or sustaining communicative competence to express needs and wants, develop social closeness, exchange information, and participate in social etiquette routines as required.

What has changed dramatically over the past 25 years, however, is how these communication goals are achieved. Whereas 25 years ago, the emphasis of AAC intervention was face to face interactions, now the scope of communication needs to be addressed has exploded to include not only face to face to interactions, but also written communication, Internet access, social media, cell phone, texting, blogging, e-commerce, etc. The expectations for the participation of individuals with complex communication needs within society also have dramatically. Whereas 25 years ago, many individuals who required AAC were living within large residential institutions with limited educational and vocational opportunities, now individuals with complex communication needs live, go to school, work, and participate within their communities (Mirenda, 2014). These changes have resulted in increased communication demands that must be addressed through AAC intervention to ensure that individuals with complex communication needs develop the necessary knowledge, judgment, and skills to ensure communicative competence.

## Knowledge, Judgment, and Skills that Support Communicative Competence

As Light (1989) first proposed, communicative competence rests on the integration of knowledge, judgment, and skills in four interrelated domains: linguistic, operational, social, and strategic. These four fundamental domains have not changed over the past 25 years. What has changed however is the breadth of linguistic, operational, social and strategic skills required to attain communicative competence. Table 1 provides a summary of the knowledge, judgment and skills required to attain communicative competence as well as examples to illustrate.

INSERT TABLE 1 HERE

**Linguistic domain.** As noted earlier, the attainment of communicative competence is predicated, at least in part, upon linguistic skills, including receptive and expressive skills in the spoken and written language(s) of the individual's home and broader social community as well as skills in the language code of the AAC systems used to communicate in these environments. The demand for linguistic skills has increased significantly over the past 25 years. As individuals with complex communication needs expand their social circles and interact with a broader audience in a wider range of environments, there are increased demands for independent, intelligible messages utilizing appropriate vocabulary, syntax, and morphology as defined by the tools and contexts of communication. There are increased demands for the development of literacy skills to facilitate access to the vast array of information technologies and social media (Williams et al., 2012). Furthermore, with increased globalization of society worldwide, more and more individuals with complex communication needs live, go to school, and work in bilingual and multilingual communities (e.g., Soto & Yu, 2014); they require receptive and expressive skills in more than one language and AAC systems to support their communication needs across different environments, thus magnifying the linguistic demands.

**Operational domain.** Beyond linguistic skills, individuals with complex communication needs also require operational skills to support communicative competence including skills in the production of unaided modes of communication and skills in access and technical operation of aided AAC systems. The need for operational skills has not changed over the past 25 years; however, with the explosion of mobile technologies and social media tools available and the current lack of universal design features across these technologies, individuals with complex communication needs face increased operational demands to effectively and efficiently access and control these diverse technologies (Emiliani et al., 2011).

**Social domain.** While linguistic and operational skills ensure that individuals with complex communication needs have access to the necessary tools to communicate, it is social skills that allow individuals with complex communication needs to use these tools effectively to meet communication goals. With the dramatic changes in the scope of communication and the media through which communication goals are attained, individuals with complex communication needs face increased demands in the social arena as well; they must learn with whom, about what, where, when, why and via what media to communicate (or not to communicate). They must learn to assess the demands of diverse audiences. With access to a much greater audience, they may face attitudinal barriers within society in many different environments (educational, vocational, social) and may need to develop increased sociorelational skills to put partners at ease and build positive relationships (Light et al., 2007; Senner, 2011).

**Strategic domain.** Despite intervention to build, re-build and /or sustain linguistic, operational, and social skills, individuals with complex communication needs will inevitably encounter situations where they face significant limitations that negatively impact their communicative competence; these situations require strategic competence. As the scope of communication needs, expectations for participation and the resulting communication demands have all increased, it is inevitable that there will be increased demands for effective coping strategies to ensure successful communication in the face of significant limitations. There is an urgent need for research to investigate strategic competence (Mirenda & Bopp, 2003); the field has much to learn from individuals who require AAC who have attained communicative competence and effectively meet their communication goals across a wide range of environments via various media (Rackensperger et al., 2005; Smith & Connolly, 2008).

## Psychosocial Factors that Support Communicative Competence

Linguistic, operational, social and strategic competence may be mitigated by a range of psychosocial factors including motivation, attitude, confidence, and resilience (Light, 2003). Table 2 provides a summary of psychosocial factors that may impact the attainment of communicative competence as well as examples to illustrate. With the increased demands of communication and the increased expectations for participation, individuals with complex communication needs will inevitably face increased communication challenges. As a result, psychosocial factors such as motivation, attitude, confidence and resilience will play an even greater role in the attainment of communicative competence than in the past. Intervention is required to foster these protective factors to ensure that individuals with complex communication needs have the drive to communicate, the willingness to use AAC, the actual propensity to do so, and the perseverance to communicate despite the many challenges and potential failures encountered (Hodge, 2007; Smith & Connolly, 2008). These issues have largely been neglected in the field to date; future research is required to advance understanding of these psychosocial factors and to improve current practices.

INSERT TABLE 2 HERE

## Environmental Supports for Communicative Competence

Since communication is a reciprocal process, communicative competence rests not just on factors related to the individual who requires AAC, but also on extrinsic factors related to the environment and communication partners (Blackstone et al., 2007). Policy, practice, attitude, skill and knowledge barriers may impede the realization of communicative competence by individuals who require AAC, whereas environmental and partner supports may serve to bolster the development, rebuilding, or maintenance of communicative competence by those with developmental, acquired, or degenerative disabilities (Beukelman & Mirenda, 2013). Environmental supports play an even greater role in the face of the increased communication challenges confronted by individuals who use AAC, especially for those who are most vulnerable. Table 3 provides a summary of environmental supports that may facilitate the development of communicative competence as well as examples to illustrate.

INSERT TABLE 3 HERE

## Future Challenges

There is no doubt that the bar has been raised. Individuals who require AAC bring a vast array of needs and skills to their communication interactions that may include significant strengths and/ or limitations in motor, sensory perceptual, cognitive, and/or language skills. The challenge is to develop effective evidence-based, culturally-competent AAC interventions to support these individuals in the realization of communicative competence to allow them to express their needs and wants, develop social closeness, exchange information, and participate in social etiquette routines as desired at home, at school, at work and/or in the community.

Twenty five years ago, the field was focused on demonstrating what was possible with access to appropriate AAC interventions (Mirenda, 2014). Now the possible is established, the challenge is to ensure that the possible becomes the probable and that every individual with complex communication needs has access to effective evidence-based AAC intervention to maximize participation and communication (Beukelman et al., 2007; Rispoli, Franco, van der Meer, Lang, & Camargo, 2010). There remain far too many individuals with complex communication needs who do not receive the effective, culturally competent, evidence-based AAC services that they require to realize communicative competence and achieve their full potential (Baxter, Enderby, Evans, & Judge, 2012; Hodge, 2007). Communicative competence is essential to the enhancement of the quality of life of individuals with complex communication needs; it is fundamental to the attainment of the basic human need, the basic human right, the basic human power of communication. As Bob Williams articulated so eloquently,

> Having the power to speak one's heart and mind changes the disability equation dramatically. In fact, it is the only thing I know that can take a sledgehammer to the age-old walls of myths and stereotypes and begin to shatter the silence that looms so large in many people's lives. (B. Williams, 2000; p. 249).

Table 1. Knowledge, judgment, and skills required for individuals who use AAC to attain communicative competence (adapted from Light & Gulens, 2002).

| Domain | Examples of knowledge, judgment, and skills required |
| --- | --- |
| Linguistic | Develop skills in the native language(s) spoken and written in the home and broader social community<br><br>• Understand the form, content, and use of spoken language(s) used by others both at home and in the broader social community<br>• Develop as many expressive skills (content, form, and use) in the spoken language(s) of the home and broader social community as appropriate<br>• Code switch between different language(s) and cultures as required<br>• Develop literacy skills to understand and use the written language(s) of the home and broader social community; code switch between these written language(s) as required<br><br>Develop skills in the language code of the AAC systems for home and the broader social community<br><br>• Develop lexical knowledge of the symbols used to express concepts via AAC<br>• Develop semantic, syntactic, and morphological skills to express more complex meanings via AAC<br>• Choose appropriate AAC systems to meet the needs of different cultural /linguistic environments<br>• Learn the appropriate linguistic conventions for different communication and social media tools |
| Operational | Produce unaided symbols. For example,<br><br>• Plan and produce the required hand shape, position, orientation, and movement to produce manual signs or conventional gestures<br>• Plan and produce the required body movements to act out messages via pantomime |

- Plan and produce the required body movements to produce other unaided codes (e.g., eye blink codes, looking up to say yes)

Operate aided AAC systems /apps accurately and efficiently. For example,

- Open communication board, turn pages, and point to target AAC symbol
- Pick up target symbol and hand it to partner when using PECS
- Use paper and pencil to draw concept
- Use selection technique to access required AAC symbols (e.g., direct selection with finger, fist, toe or eyes; row column scanning with a single switch; directed scanning with a joystick)
- Navigate within AAC systems/ apps as required

Operate social media and other mainstream communication tools

- Access social media /communication tools as required
- Capture and upload photos and video as required to support communication via social media
- Navigate between apps/ tools as required to meet needs

Social      Develop appropriate sociolinguistic skills

- Fulfill obligatory and nonobligatory turns in interaction
- Initiate and terminate interactions appropriately
- Maintain and develop topics of conversation
- Express a wide range of communicative functions (e.g., request information, protest, request objects/actions, provide information, provide clarification, confirm/deny, request attention)
- Choose appropriate AAC systems /apps and/ or social media tools to meet communication needs as required
- Use appropriate form, content, and conventions as required for the audience and media

Develop appropriate sociorelational skills

- Participate actively in interactions
- Be responsive to partners
- Demonstrate interest in partners (e.g., ask partner-focused questions)
- Put partners at ease
- Project a positive self-image
- Maintain a positive rapport with partners

Strategic      Use compensatory strategies to bypass limitations in the linguistic domain. For example,

- Ask partner to write /type or point to symbols to augment spoken input and bypass comprehension difficulties
- Use mementos to bypass vocabulary limitations and establish the topic
- Ask partner to provide choices to overcome vocabulary limitations
- Ask the partner to guess and provide clues to bypass vocabulary limitations

Use compensatory strategies to bypass limitations in the operational domain. For example,

- Use telegraphic messages to enhance rate of communication
- Ask partner to predict as message is spelled to reduce fatigue and enhance rate of communication
- Have partner assist in locating appropriate page to help with navigational demands

Use compensatory strategies to bypass limitations in social domain

- Use an introduction strategy to put the partner at ease
- Use humor to maintain a positive rapport and put partner at ease
- Utilize social media to increase social network

Table 2. Psychosocial factors and the potential impact on communicative competence (adapted from Light, 2003)

| Psychosocial factor | Definition | Potential impact |
| --- | --- | --- |
| Motivation to communicate | Drive to communicate, influenced by the belief that the goal (i.e., communication) is important and attainable | Defines the individual's desire to communicate with others in specific situations |
| Attitude toward AAC | Ideas about AAC charged with emotion (positive or negative) that predispose AAC use (or lack of use) in a given situation | Influences the individual's willingness to use (or not use) AAC to communicate with others in specific situations |
| Communication confidence | Self-assurance based on the individual's belief that communication success is achievable within a given situation | Influences the propensity of the individual to actually act (i.e., communicate) in specific situations |
| Resilience | Capacity to prevent, minimize, or overcome the damaging effects of adversities; capacity to compensate for problems and recover from failures | Influences the individual's persistence with communication in the face of barriers, adversities, and failures |

Table 3. Environmental supports that may facilitate the communicative competence of individuals who require AAC (adapted from Light, 2003)

| Environmental factor | Examples of environmental supports |
|---|---|
| Policy | • Legislation that supports accessibility and inclusion of individuals who require AAC<br>• Policies that ensure funding of AAC systems and assistive technologies<br>• Legislation that prohibits discrimination against individuals with disabilities<br>• Policies that support universal design of technologies |
| Practice | • Evidence-based, consumer responsive, culturally competent service delivery by multidisciplinary team with expertise in AAC<br>• Funding support for AAC systems/ assistive technologies and services<br>• Availability of technologies that are accessible for individuals with disabilities |
| Attitude | • Advocacy and public education activities to promote awareness of the rights and capabilities of individuals with disabilities<br>• Meaningful opportunities for communication and interaction with peers<br>• Appropriate expectations in the home, school, work and community |
| Knowledge | • Knowledge of funding sources and AAC resources<br>• Knowledge of AAC symbols and transmission techniques<br>• Knowledge of positioning requirements<br>• Knowledge of strategies for vocabulary selection, layout, organization, and regular updating |

- Knowledge of operation and programming of AAC technologies
- Knowledge of daily care and maintenance routines
- Strategies for technical trouble shooting
- Strategies for integrating AAC into daily use

Skills
- Partners who use interaction strategies to support successful communication (e.g., wait for individual to communicate, recognize and respond to communicative attempts, provide appropriate language input, augment input if required, confirm their understanding)